# dbMPIKT: A web resource for the kinetic and thermodynamic database of mutant protein interactions

Quanya Liu [1], Peng Chen [1,2] *, Bing Wang [4] and Jinyan Li [3,*]

[1]Institute of Health Sciences, Anhui University, 230601 Hefei, Anhui, China. [2]School of Computer Science and Technology, Anhui University, 230601 Hefei, Anhui, China. [3]Advanced Analytics Institute and Centre for Health Technologies, University of Technology, Sydney, Broadway, NSW, 2007 Australia. [4]School of Electrical and Information Engineering, Anhui University of Technology, 243032 Ma'anshan, Anhui, China.

## ABSTRACT

Protein-protein interactions (PPIs) perform important roles on biological functions. Researches of mutants on protein interactions can further understand PPIs. In the past, many researchers have developed databases that stored mutants on protein interactions, which are old and not updated till now. To address the issue, we developed a kinetic and thermodynamic database of mutant protein interactions (dbMPIKT) that can be freely accessible at our website. This database contains 5291 mutants that integrated data from previous databases and data from literatures for nearly three years. Furthermore, the data were analyzed, involving mutation number, mutation type, protein pair source and network map construction. On the whole, the database provides new data to further improve the study on PPIs. Website: http://210.45.212.128/lqy/index.php

Keywords: PPIs; mutants; protein interaction; kinetic data; thermodynamic data.

## INTRODUCTION

Protein-protein interactions (PPIs) are involved in all aspects of the organism in life. PPIs play crucial roles in mediating the majority of biological functions (1), which are also associated with human diseases. Mutations in PPIs always cause some diseases, for instance, cancer and Alzheimers disease (2). By studying PPIs, the mechanism of interactions can be further studied and can be used to treat intervention and drug design for more researches (3). The understanding of PPIs can be used to design PPI inhibitors (4). As we all known that one PPI interface contains many amino acid residues, in which only a small number of residues contribute major binding free energy, which are defined as hot spots (5). Hot spots play crucial roles in regulating PPIs, while the knowledge of hot spots is extremely important in designing PPI inhibitors (6). Many researchers have developed different methods to obtain the mutants of protein interactions information in previous literatures, and used these data to establish public databases that can be more convenient for users to study on hot spots (7). These databases provided a useful platform for integrating a lot of data on hot spots directly.

Hot spots can be calculated by some formulas on the mutant data of protein interactions. More and more researchers have focused on hot spots and have built up several databases of mutants data associated with hot spots. The first database of alanine mutations in protein interactions was built by Thorn and Bogan (8). It is called ASEdb that collected experimentally determined binding affinity data. The protein interactions refer to the interactions between proteins and other biological macro-molecules, such as proteins, nucleic acids and small molecules. Then, BID was developed by Fischer *et al* that mining hot spots in protein interfaces, in which the protein interaction information was extracted from scientific literatures. With the help of tables, visualization tools and supporting documents, users can get detailed protein interaction information in BID (9). Then, PINT was formed by Kumar and Gromiha. The PINT mainly shows the thermodynamic data on PPIs, such as binding free energy change, dissociation constant, heat capacity change and so on. This database presents the protein names, mutation residues, literature information and measuring methods of PPIs (10). Next, SKEMPI was a manually curated database that containing 3046 binding free energy changes upon mutation in literatures (11). In the above mentioned database, many interactive databases can provide the mutant information of protein interactions.

Study of the mutants of protein interactions are always based on experimental data, which can be used to further improve the study of hot spots. Hot spots can be found using biological experiments such as alanine scanning mutagenesis and Alanine shaving (12). The mostly accepted definition of hot spots is that residues having alanine mutation with changes in binding free energy ($\Delta\Delta G$) $\geq$ 2.0 kcal/mol are hot spots (HS), others non-hot spots (NS) (13). To experimentally identify hot spots, a series of experiments have developed, including the mutation of all amino acid residues at PPI interfaces as well as the purification and expression of mutant proteins. The most important step is to measure their binding affinities (14). Experimental methods have many drawbacks including not only time-consuming and labor-intensive, but also unable to measure all potential binding hot spots in a large number of proteins (15). In order to improve research efficiency, many researchers applied computational tools to identify hot spots. Machine learning methods were most widely used in the identification of hot spots (16). The commonly used machine learning methods were SVM, MM-PBSA and MM-GBSA (17). They used the established related database as training set to build training model and then applied it to the predictions on unknown amino acid residues (18). But in recent years, these databases did not be updated in time. To address the issue, we constructed a new database that mining the mutants of protein interactions from related database data and literatures.

In this article, we present a kinetic and thermodynamic database of mutant protein interactions, called dbMPIKT. The dbMPIKT contains 5291 mutants of experimentally kinetic and thermodynamics data upon mutation. The database consists of the data from previous databases about mutant protein interactions, including BID, SKEMPI and AB-Bind, and the data extracted from scientific literatures in recent years. From this database, researchers can expand their researches for predicting hot spots, drug discovery and development, and other relevant researches. Next, we also make statistical analysis for these data and construct a hot spot interaction network, which can be promise in drug design (19).

## MATERIALS AND METHODS

### Data collection

This database consists of four data sources, SKEMPI, BID, AB-Bind and dbMPIKT. The first three data sources are derived from other databases that have been built before. The dbMPIKT data collected the mutant data of mutant protein interactions from scientific literatures for nearly three years. To build the complete database of hot spots, the first step is to integrate the first three data sets and remove redundancy. This is a part of our database data, and the other part is dbMPIKT data that derived from scientific literatures. Firstly, to build dbMPIKT, we performed a comprehensive literature search in NCBI to find related literatures in PubMed. As a result, 425 literatures were obtained by means of different keywords. Then, the kinetic and thermodynamics data of mutants were read and extracted from these literatures. However, we found that many related literatures are missed. Secondly, we searched for protein complex structures in the PDB databank. Some protein structures were filtered out under some conditions, such as resolution of $3Å$ and more, only protein structures (heterodimer), the time limit is from 1 January 2013 to 31 December 2016. Afterwards, 1017 protein structures were obtained from 682 citations. Then these structures were mapped to PDB-Bind and searched for their corresponding thermodynamic data. As a result, 99 complex structures from 85 citations were found containing dissociation constant ($K_d$ value). The most important step is that we read these literatures and recorded $K_d$ values of these structures (20). Finally, our database composed of the manually curated data and the other three data sources.

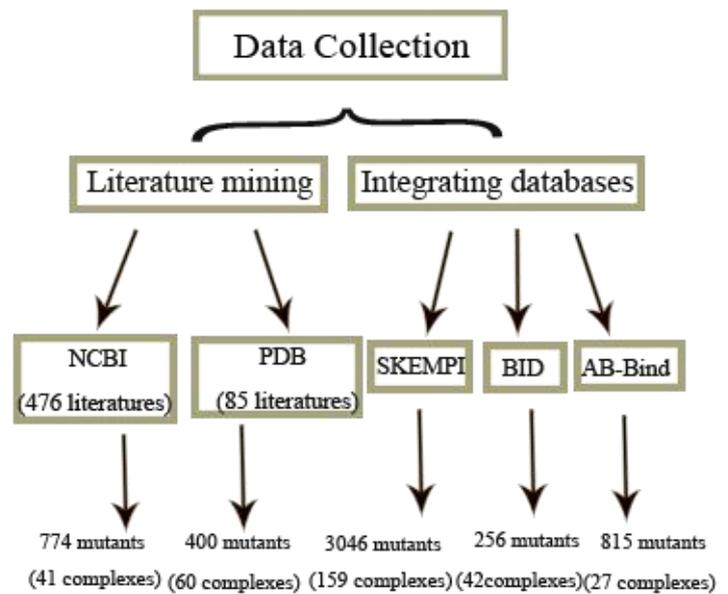

**Figure 1.** The flow chart of data collection.

### Database construction

The structure of the database was implemented with MySQL database and all functions of the webserver were implemented in PHP script. The webserver consists of home page and the pages of browse, document, upload, download and contact us. Figure 2 shows the database structure. More information about the database can be seen by browsing the six web pages.

In the home page, the database is briefly introduced and some statistics in the database are simply presented. In addition, a quick search on the right part of the page was set up. By typing one PDB ID, users can search for the protein you want in the database and get relevant mutant information, or if the query protein is not listed in the database, no information will be returned. In the bottom right corner of the page, some links are shown for transferring to other related sites.

In the Browse page, we presents all data in the database. Here you can see details of mutants for the four sources. All data can be freely downloaded from the webserver. If you want to download all data, you can click on the page of "Download". In order to continuously update the database, we provide a "upload" link that help users uploading their own data, checking them and storing into the database through the upload interface. In addition, the newly uploaded data will also be placed on the "browse" page. If you have any question, please click on "contact us" and you can contact us by an email.

### Analysis of protein-protein interaction pairs and Interaction network construction

In the process of mutation data mining, protein-protein interaction pairs were also recorded. According to PDB ID, the source of each pair of PPI was founded from the PDB database. We classified all protein-protein complexes into different categories that are based on atomic structures of complexes. In order to illustrate whether each pair of PPI is linked, the network analysis tool (Cytoscape version 3.5.1) [13]. was used to build the interaction network. The regularity of PPIs were obtained by analyzing the association of PPIs, and some conclusions were also made according to the obtained network map.

## RESULTS

### Data entry in database

In this article, a webserver was built for kinetic and thermodynamic database of mutant protein interactions. We present all data in four tables. Although data entries are not all consistent, they include the following eight attributes. The first one is PDB ID, which is the ID of protein-protein complex in PDB database. It links to related PDB website, so you can get more information of the complex by clicking it. Second, mutations are important data attributes. The description of mutations includes original residue, chain identifier, the number of residues representing the mutation position and mutant residue. Third, protein 1 and protein 2 denote the names of the two proteins that are interacted. The next attributes are kinetic data and thermodynamic data. In general, kinetic data contains ($K_d$), association rate ($K_{on}$), dissociation rate ($K_{off}$) and so on. Most of them are shown in units of $nM$, $M^{-1}S^{-1}$ and $S^{-1}$. Other units can be converted to them. Moreover, thermodynamic data contains change in binding free energy ($\Delta G$) and difference in binding free energy change between mutant and wild-type complex ($\Delta\Delta G$). They were reported in the unit of *kcal/mol*. Next attribute is PubMed ID, which is the source of kinetic and thermodynamic data. In addition, you can refer to more details by clicking on "PubMed ID" in table and downloading literature from NCBI. Furthermore, the last attribute is "method", which presents the experimental measurement method of the affinity. There are mainly two methods: SPR (Surface Plasmon Resonance) and ITC (isothermal titration calorimetric) (22). In addition, temperature information is described in the attribute.

### Database statistics

To learn more information on mutant protein interactions, statistical tools were used to analyze our database data. In this database, we collected 5291 mutants having kinetic or thermodynamic data. Then these data were divided into four groups based on their sources: SKEMPI, AB-Bind, BID and dbMPIKT. Among them, SKEMPI, AB-Bind, BID and dbMPIKT contain 3046, 815, 256 and 1174 mutants, respectively. These mutants are derived from 233 structures of 245 protein-protein complexes, only 12 complexes do not have PDB ID. On the one hand, we compared the four datasets with respect to the type of mutations. The collected data in the database is clustered into three mutation types: single mutants, double mutants and multiple mutants. The data sources and data distribution of the database are shown in figure 3 (The table S1 can be referred to the supplementary materials). In general, SKEMPI data contains the most number of single mutants and dbMPIKT data contains the second most mutation type. On the other hand, for the mutation type, single mutations account for 75.88% of the total mutations, double mutations are 13.28% and multiple mutations are 10.84%. Specifically, the collected thermodynamic and kinetic data are measured by $\Delta\Delta G$, change in enthalpy ($\Delta H$), change in entropy ($\Delta S$), and kinetic rate constants and so on, which comes from the experiments of SPR, ITC and Alanine mutation scanning (AMS).

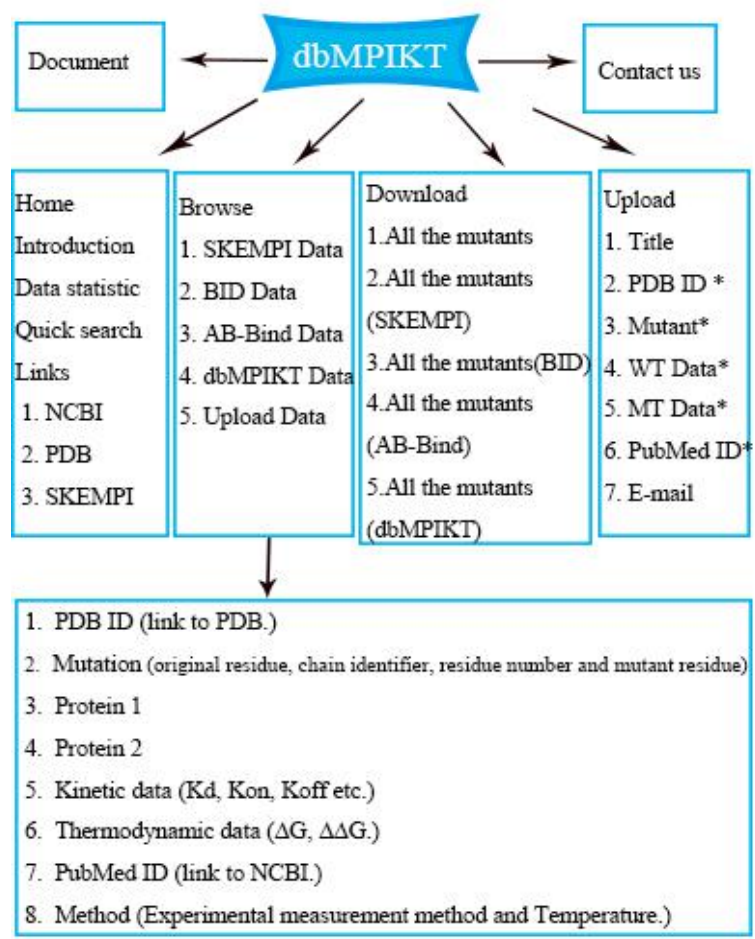

**Figure 2.** The database structure of dbMPIKT.

The database stored almost all experimental mutants up to now. For single mutations, we counted the number of mutations for each type of amino acids. Table 1 shows the description of 20 amino acids on the single mutant data. Statistically, alanine mutation accounts for 56% in the single mutant data, while the other amino acid mutations are 44% in which threonine is the least type. Considering different amino acid properties, the 20 types of amino acids are divided into five categories: polar (S, T, N and Q), hydrophobic (A, I, L, M, V, W, Y and F), positive (R, K and H), negative (D and E) or other (G, P and C) (23). In comparison with other data sets, the number of alanine mutations in dbMPIKT is the largest in the four data sets. In dbMPIKT data set, alanine mutations are accounted for 66.7%, and in general, the number of alanine mutations is also the most type in our database.

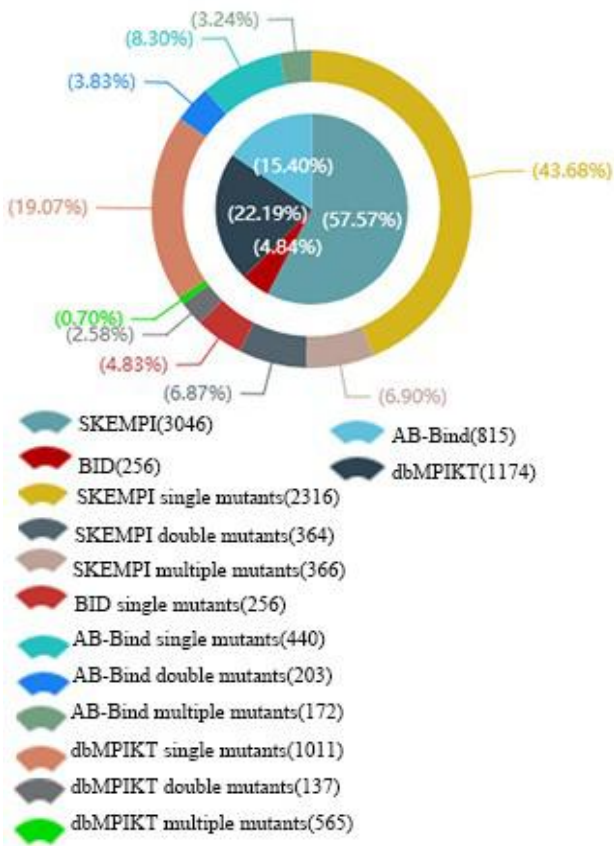

**Figure 3.** The data distribution of the four data sets.

## Analysis of protein-protein pairs in dbMPIKT

In our database, there are 5291 mutants. During the collection of mutant information, the corresponding protein interactions were also recorded. All mutants come from 245 protein-protein complexes that representing the 245 protein-protein interaction pairs. In our database, the protein-protein complexes are heterodimer complexes, antigen-antibodies, enzyme-inhibitors and so on (24), which are all binary. The proteins are human proteins, mus musculus proteins, bos taurus proteins and so on. Among them, human proteins are the most type. Based on the protein pairs, protein interaction network was constructed. Figure 4 illustrates a part of the protein interaction network, while the complete network can be seen in figure S2 of the supplementary material. In Figure S2, most of the protein interactions are independent, whereas a small portion of proteins interact with each other to form an interaction network (see Figure 4). Protein interaction network can be used to identify protein functions for specific protein interactions (25). In Figure 4, we can see a small network which is centered at basic pancreatic trypsin inhibitor (BPTI) and Bovine alpha-chymotrypsin protein, which are bos taurus proteins. BPTI is a globular protein that can be used to study conformation and PPIs, which can also play an important role in Biomedical Science. In Clinical, it can be used to reduce hemorrhagic complications (26). In addition, researches can use the protein interaction network to analyze protein biological functions (27).

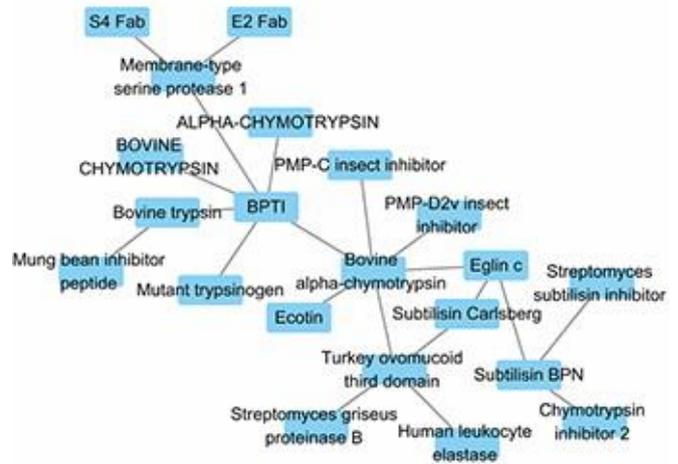

**Figure 4.** Part of the interaction on the network map. Each node in the picture represents a protein, and the connection between nodes represents an interaction.

## DISCUSSION

In this article, we have four data sets. They all include kinetic or thermodynamic data of mutant protein interactions. But they are somewhat different. In general, the SKEMPI database contains the largest number of mutants, followed by the manually curated dbMPIKT. Since the website of BID database doesnot work now, we cannot download data from it and we get some BID data from the additional material of a literature (28). Therefore the BID is the least database used here. In addition, in the case of mutant types, the dbMPIKT contains the largest alanine mutations. It suggests that the data of our database is more useful for the hot spot predictions. Moreover, based on protein types, previous databases are almost targeted for specific complexes. For example, AB-bind is an antibody binding mutational database extracted information from antigen-antibody complexes. The paper proposes to integrate these databases together so that researchers can be easy to get required data.

**Table 1.** Kinds and the number of single mutant amino acids statistics

| Datastes | Amino | SKEM | BI | AB- | dbMPI | Tot |
|---|---|---|---|---|---|---|
| Hydrophob | A | 1057 | 2 | 2 | 6 | 22 |
| Hydrophob | I | 5 | 0 | 2 | 1 | 6 |
| Hydrophob | L | 7 | 0 | 3 | 1 | 9 |
| Hydrophob | M | 6 | 0 | 3 | 1 | 7 |
| Hydrophob | V | 7 | 0 | 1 | 1 | 9 |
| Hydrophob | W | 5 | 0 | 1 | 1 | 8 |
| Hydrophob | Y | 5 | 0 | 1 | 1 | 8 |
| Hydrophob | F | 9 | 0 | 8 | 2 | 12 |
| Hydrophob | S | 7 | 0 | 1 | 2 | 10 |
| P | T | 4 | 0 | 6 | 9 | 6 |
| P | N | 6 | 0 | 1 | 2 | 9 |
| P | Q | 7 | 0 | 2 | 2 | 12 |
| Positive | R | 7 | 0 | 1 | 1 | 10 |
| Positive | K | 9 | 0 | 7 | 2 | 13 |
| Positive | H | 5 | 0 | 2 | 1 | 6 |
| Negative | D | 7 | 0 | 1 | 2 | 11 |
| Negative | E | 8 | 0 | 1 | 3 | 13 |
| Ot | C | 4 | 0 | 1 | 1 | 6 |
| Ot | G | 5 | 0 | 7 | 2 | 8 |
| Ot | P | 7 | 0 | 2 | 2 | 12 |
| Total | | 2316 | 256 | 440 | 1010 | 4022 |

## CONCLUSION

The paper integrated the previous three databases and manually collected data from literatures in nearly last three years. We built a web server to store kinetic and thermodynamic data of mutant protein interactions. More detailed information about mutants and protein-protein interactions is presented in the database. In our database, the kinetic or thermodynamic data of mutants, such as, $K_d$, $\Delta\Delta G$, $\Delta G$, $K_{off}$ and $K_{on}$, are obtained. In addition, some data can be calculated by others, for examples,

$$K_d = \frac{K_{on}}{K_{off}} \quad \text{and} \quad \Delta G = RT \ln K_d \quad (1)$$

We can get $\Delta\Delta G$ indirectly by formula calculation, which can diametrically distinguish between hot spots and non-hot spots. The database provides a large hot spot data set that can help improve the applications of hot spots and the hot spot predictions.

## ACKNOWLEDGEMENTS

This work was supported by the National Natural Science Foundation of China (Nos. 61672035, 61300058 and 61472282).

## COMPETING INTERESTS

The authors declare that they have no competing interests.

## AUTHOR'S CONTRIBUTIONS

QL and PC conceived the study; QL and BW participated in the database design; QL, CZ and PC carried it out and drafted the manuscript. All authors revised the manuscript critically. JL and PC approved the final manuscript.

## REFERENCES

1. N London, B Raveh, and O Schueler-Furman. (2013) DruggableProtein-Protein Interactions–from Hot Spots to Hot Segments. *Current Opinion in Chemical Biology*, **17**, 952.
2. G. Hu, F. Xiao, Y. Li, Y. Li, and W Vongsangnak. (2016) Protein-Protein Interface and Disease: Perspective from Biomolecular Networks. *Advances in Biochemical Engineering/biotechnology*.
3. Leonardo G. Ferreira, Glaucius Oliva, and Adriano D. Andricopulo. (2016) Protein-Protein Interaction Inhibitors: Advances in Anticancer Drug Design. *Expert Opinion on Drug Discovery*, **11**, 957.
4. O Zarei, M Hamzehmivehroud, S Benvenuti, F Ustunalkan, and S Dastmalchi. (2017) Characterizing the Hot Spots Involved in Ron-Msp Complex Formation Using in Silico Alanine Scanning Mutagenesis and Molecular Dynamics Simulation. *Advanced Pharmaceutical Bulletin*, **7**, 141-50.
5. Cristian R. Munteanu, Antnio C. Pimenta, Carlos Fernandez-Lozano, Andr Melo, Maria N. D. S. Cordeiro, and Irina S. Moreira. (2015) Solvent Accessible Surface Area-Based Hot-Spot Detection Methods for Protein-Protein and Protein-Nucleic Acid Interfaces. *Journal of Chemical Information & Modeling*, **55**, 1077.
- Logan R. Hoggard, Yongqiang Zhang, Min Zhang, Vanja Panic, John A. Wisniewski, and Haitao Ji. (2015) Rational Design of Selective Small-Molecule Inhibitors for -Catenin/B-Cell Lymphoma 9 ProteinCProtein Interactions. *Advanced Pharmaceutical Bulletin*, **137**, 12249-60.
- Samuel Kerrien, Bruno Aranda, Lionel Breuza, Alan Bridge, Fiona Broackescarter, Carol Chen, Margaret Duesbury, Marine Dumousseau, Marc Feuermann, and Ursula Hinz. (2012) The Intact Molecular Interaction Database in 2012. *Nucleic Acids Research*, **40**, D841.
- K. S. Thorn, and A. A. Bogan. (2001) Asedb: A Database of Alanine Mutations and Their Effects on the Free Energy of Binding in Protein Interactions. *Bioinformatics*, **17**, 284.
- T. B. Fischer, K. V. Arunachalam, D. Bailey, V. Mangual, S. Bakhru, R. Russo, D. Huang, M. Paczkowski, V. Lalchandani, and C. Ramachandra. (2003) The Binding Interface Database (Bid): A Compilation of Amino Acid Hot Spots in Protein Interfaces. *Bioinformatics*, **19**, 1453.
- M. D. Shaji Kumar, and M. Michael Gromiha. (2017) Pint: ProteinCProtein Interactions Thermodynamic Database. *Nucleic Acids Research*, **34**, D195.
- I. H. Moal, and J Fernndezrecio. (2012) Skempi: A Structural Kinetic and Energetic Database of Mutant Protein Interactions and Its Use in Empirical Models. *Bioinformatics*, **28**, 2600.
- I. S. Moreira, J. M. Martins, R. M. Ramos, P. A. Fernandes, and M. J. Ramos. (2013) Understanding the Importance of the Aromatic Amino-Acid Residues as Hot-Spots. *Biochimica Et Biophysica Acta*, **1834**, 404-14.
(21) Bin Xu, Xiaoming Wei, Deng Lei, Jihong Guan, and Shuigeng Zhou. (2012) A Semi-Supervised Boosting Svm for Predicting Hot Spots at Protein-Protein Interfaces. *Bmc Systems Biology*, **6 Suppl 2**, S6.
(22) J Shirian, O Sharabi, and J. M. Shifman. (2016) Cold Spots in Protein Binding. *Trends in Biochemical Sciences*, **41**, 739.
(23) Liu Qian, Ren Jing, Jiangning Song, and Jinyan Li. (2015) Co-Occurring Atomic Contacts for the Characterization of Protein Binding Hot Spots. *Plos One*, **10**.
(24) Rita Melo, Robert Fieldhouse, Andr Melo, Jo?o D. G. Correia, Maria Natlia D. S. Cordeiro, Zeynep H. Gm?, Joaquim Costa, Alexandre M. J. J. Bonvin, and Irina S. Moreira. (2016) A Machine Learning Approach for Hot-Spot Detection at Protein-Protein Interfaces. *International Journal of Molecular Sciences*, **17**, 1215.
(25) S Monta?o, E Orozco, J Correa-Basurto, M Bello, B Chvez-Munguła, and A Betanzos. (2017) Heterodimerization of the Entamoeba Histolytica Ehcpadh Virulence Complex through Molecular Dynamics and Protein- Protein Docking. *Journal of Biomolecular Structure & Dynamics*, 486- 503.
(26) J. R. Brender, and Y. Zhang. (2015) Predicting the Effect of Mutations on Protein-Protein Binding Interactions through Structure-Based Interface Profiles. *Plos Computational Biology*, **11**, e1004494.
(27) R. Agius, M. Torchala, I. H. Moal, J. Fernandez-Recio, and P. A. Bates. (2013) Characterizing Changes in the Rate of Protein-Protein Dissociation Upon Interface Mutation Using Hotspot Energy and Organization. *PLoS Comput Biol*, **9**, e1003216.
(28) J. Chen, N Sawyer, and L Regan. (2013) Protein-Protein Interactions: General Trends in the Relationship between Binding Affinity and Interfacial Buried Surface Area. *Protein Science*, **22**, 510.
(29) Gang Su, John H Morris, Barry Demchak, and Gary D Bader. (2014) Biological Network Exploration with Cytoscape. *Current Protocols in Bioinformatics*, **47**, 8.13.1.
(30) Current Protocols in Bioinformatics. (2012) Mutational Properties of Amino Acid Residues: Implications for Evolvability of Phosphorylatable Residues. *Philosophical Transactions of the Royal Society of London*, **367**, 2584.
(31) K. S. M. Tozammel Hossain, Chris Bailey-Kellogg, Alan M. Friedman, Michael J. Bradley, Nathan Baker, and Naren Ramakrishnan. (2011) Using Physicochemical Properties of Amino Acids to Induce Graphical Models of Residue Couplings. *in Tenth International Workshop on Data Mining in Bioinformatics*, pp. 1-10.
(32) R. P. Bahadur, and M. Zacharias. (2008) The Interface of Protein-Protein Complexes: Analysis of Contacts and Prediction of Interactions. *Cell Mol Life Sci*, **65**, 1059-72.
(33) Christine Brun, Fran?ois Chevenet, David Martin, Jr?me Wojcik, Alain Gunoche, and Bernard Jacq. (2003) Functional Classification of Proteins for the Prediction of Cellular Function from a Protein-Protein Interaction Network. *Genome Biology*, **5**, R6.
(34) P Ascenzi, A Bocedi, M Bolognesi, A Spallarossa, M Coletta, Cristofaro R De, and E Menegatti. (2003) The Bovine Basic Pancreatic Trypsin

Inhibitor (Kunitz Inhibitor): A Milestone Protein. *Current Protein & Peptide Science*, **4**, 231.

(35)    A Vazquez, A Flammini, A Maritan, and A Vespignani. (2003) A Global Protein Function Prediction in Protein-Protein Interaction Networks. *Nature Biotechnology*, **21**, 697-700.

(36)    Nurcan Tuncbag, Attila Gursoy, and Ozlem Keskin. (2013) Identification of Computational Hot Spots in Protein Interfaces. *Oxford University Press*, p. 1513.

# ADDITION FILE

Additional file 1

Figure S1 Mutation type analysis.

Additional file 2

Figure S2 Protein interaction network map.

Additional file 3

Table S3 Data sources and data distribution.

Additional file 4

Table S4 The collection of protein complexes and their sources.

Additional file 5

Table S5 Literature for the collected data.